\documentclass[review]{elsarticle}

\journal{Journal of \LaTeX\ Templates}

\begin{document}

\begin{frontmatter}

\title{p-process chaser detector in $n$ - $\gamma$ coincidences}


\author[SARI,Konan]{H. Utsunomiya\corref{CA}}
\cortext[CA]{Corresponding author}
\ead{hiro@sari.ac.cn}
\author[SARI,BEIJING]{Z.R. Hao}
\author[ULB]{S. Goriely}
\author[SARI,BEIJING,SINAP]{X.G. Cao}
\author[SARI,BEIJING,SINAP]{G.T. Fan}
\author[SARI,BEIJING,SINAP]{H.W. Wang}

\address[SARI]{Shanghai Advanced Research Institute, Chinese Academy of Sciences, Shanghai 201210, China}
\address[Konan]{Department of Physics, Konan University, Kobe 658-8501, Japan}
\address[BEIJING]{University of Chinese Academy of Sciences, Beijing 100049, China}
\address[ULB]{Institut d'Astronomie et d'Astrophysique, Universit\`e Libre de Bruxelles, 1050 Brussels, Belgium}
\address[SINAP]{Shanghai Institute of Applied Physics, Chinese Academy of Sciences, Shanghai 201800, China}

\begin{abstract}
We propose two types of neutron-$\gamma_1$-$\gamma_2$ triple coincidence detectors (not constructed) to chase gamma transitions to produce p-nuclei following the neutron emission in the $(\gamma, n)$ reaction.  Neutrons are detected with 24 $^3$He counters embedded in a polyethylene moderator in Type I detector and with 6 liquid scintillation detectors in Type II detector, respectively. $\gamma$ rays are detected with two high-purity germanium detectors and four LaBr$_3$(Ce) detectors.
The detector which is referred to as p-process chaser detector is used to search for mediating states in $^{180}$Ta through which the isomeric and ground states in $^{180}$Ta are thermalized in the p-process. A search is made for both resonant states and unresolved states in high nuclear-level-density domain. 
\end{abstract}

\begin{keyword}
\texttt{neutron - $\gamma$ coincidence detector} \sep $^3$He counter \sep neutron moderator \sep liquid scintillation detector \sep HPGe detector \sep LaBr$_3$(Ce) detector \sep p-process study of $^{180}$Ta 
\end{keyword}

\end{frontmatter}


\section{Introduction}
Photodisintegration plays the leading role in the p-process \cite{ArGo03}. The most promising astrophysical site is the deep O-Ne-rich layer of massive stars exploding as type II supernovae \cite{WH78,Ray95,Rau02} which satisfactorily meets the temperature condition to facilitate photodisintegrations above T$_9$=1.5 on the one hand and to prevent the photoerosion below T$_9$=3.5 on the other hand, and the time condition to freeze out photodisintegrations on a short time-scale \cite{ArGo03,Utsu06}. The production of heavy neutron-deficient nuclei so called p-nuclei is triggered by photodisintegrating seeds nuclei that are produced in the slow and/or rapid neutron capture processes and pre-existing in the deep layers of massive stars.  Another astrophysical site that may contribute to the Galactic enrichment of p-nuclei concerns type Ia supernovae corresponding to the disruption of a white dwarf after having accreted material above the Chandrasekhar limit from its companion in a binary system \cite{Trav15}.

The $^{180}$Ta is classified as a p-nucleus \cite{Ande89,Palm93} since it is expected to be mainly produced by the p-process \cite{ArGo03}. It exists as the only naturally occurring isomer at the excitation energy E$_x$ = 77.2 keV with the spin-parity $J^{\pi}$ = 9$^-$ and half-life $T_{1/2}$ in excess of 7.1 $\times$ 10$^{15}$ yr \cite{NNDC} and rarest element in the solar system. The total $^{181}$Ta($\gamma$,n)$^{180}$Ta cross section of direct relevance to the production of $^{180}$Ta was measured \cite{Utsu03}. The survival of $^{180}$Ta relies on the thermalization between the isomeric state $^{180}$Ta$^m$ and ground state $^{180}$Ta$^{gs}$ with $J^{\pi}$ = 1$^+$ and $T_{1/2}$ = 8.15 hr through mediating states under the stellar condition. Although the $\gamma$-spectroscopy experiment apparently did not show the presence of mediating states that links the 9$^-$ and 1$^+$ states by $\gamma$ transitions \cite{Drac98}, the photoactivation experiment with bremsstrahlung showed an acceleration of the destruction of the isomeric states through the ground state at $E_x$ $>$ 1 MeV \cite{Beli99}, suggesting that the two states,  $^{180}$Ta$^m$ and $^{180}$Ta$^{gs}$, are thermalized at stellar temperatures above $T_9$ $\geq$ 0.4 \cite{ArGo03}.

Experimental search for the mediating states is a difficult task as shown by the $\gamma$-spectroscopy experiment \cite{Drac98}. The partial $^{181}$Ta($\gamma$,n)$^{180}$Ta$^m$ cross section measurement \cite{Goko06} showed that $\gamma$ transitions to the isomeric states are favored by high-spin states with $J$ $>$ 5 in the relevant excitation energy region. This nature of $\gamma$ transitions via high-spin states to the isomeric states is also expected in the s-process production of $^{180}$Ta$^m$ through $^{179}$Hf($\beta$)$^{179}$Ta(n,$\gamma$)$^{180}$Ta$^m$ \cite{YoTa83}. Since the nuclear level density (NLD) of the odd-odd nucleus $^{180}$Ta is very high \cite{Hila01} reaching $\sim$ 10$^5$ levels per MeV at the excitation energy of 6 MeV \cite{Gori08}, mediating states are not necessarily discrete resonant states, but could be unresolved states in the high NLD domain.    

We propose to search for the mediating states by chasing the $^{180}$Ta production along the p-process path, $^{181}$Ta($\gamma$,n)$^{180}$Ta.  In this paper, we report on two types of neutron-$\gamma_1$-$\gamma_2$ triple coincidence detectors (not constructed) for the chase study that we refer to as p-process chaser detector.  

\section{p-process chaser detector}
\subsection{Detector concept}
The $^{180}$Ta$^m$ and $^{180}$Ta$^{gs}$ are produced by $\gamma$ transitions from excited states in $^{180}$Ta that are populated by the neutron emission in the $^{181}$Ta($\gamma$,n)$^{180}$Ta reaction. Figure 1 depicts the production process. $\gamma$-ray energies relevant to the p-process lie in the low-energy tail of the giant-dipole resonance \cite{Utsu06}. The chase study is performed below particle threshold in $^{180}$Ta, the proton threshold at 5.7 MeV which corresponds to the $\gamma$-ray energy at 13.3 MeV. In this energy region, only ($\gamma$,n) reactions on $^{181}$Ta occur. Previously, the population of excited states in $^{180}$Ta was investigated in ($\gamma$,n) reactions induced by monochromatic $\gamma$ rays from nuclei produced by thermal neutron capture in the reactor \cite{Cous84}. One can investigate the excited state population and neutron branching ratios by measuring neutrons from the  ($\gamma$,n) reaction with the time-of-flight (TOF) technique and measure $\gamma$ transitions to the isomeric and ground states in $\gamma$-$\gamma$ coincidences. It is ideal to perform the neutron TOF measurement and $\gamma$-$\gamma$ coincidence measurement in one experiment. However, such a triple coincidence measurement involving a high-resolution TOF measurement may not be feasible from the point of view of the overall detection efficiency. We will not discuss such a separate high-resolution neutron TOF measurement in this paper.

\begin{figure}
\center
\vskip -2cm
\includegraphics[bb = 270 100 700 450, width = 110 mm, clip]{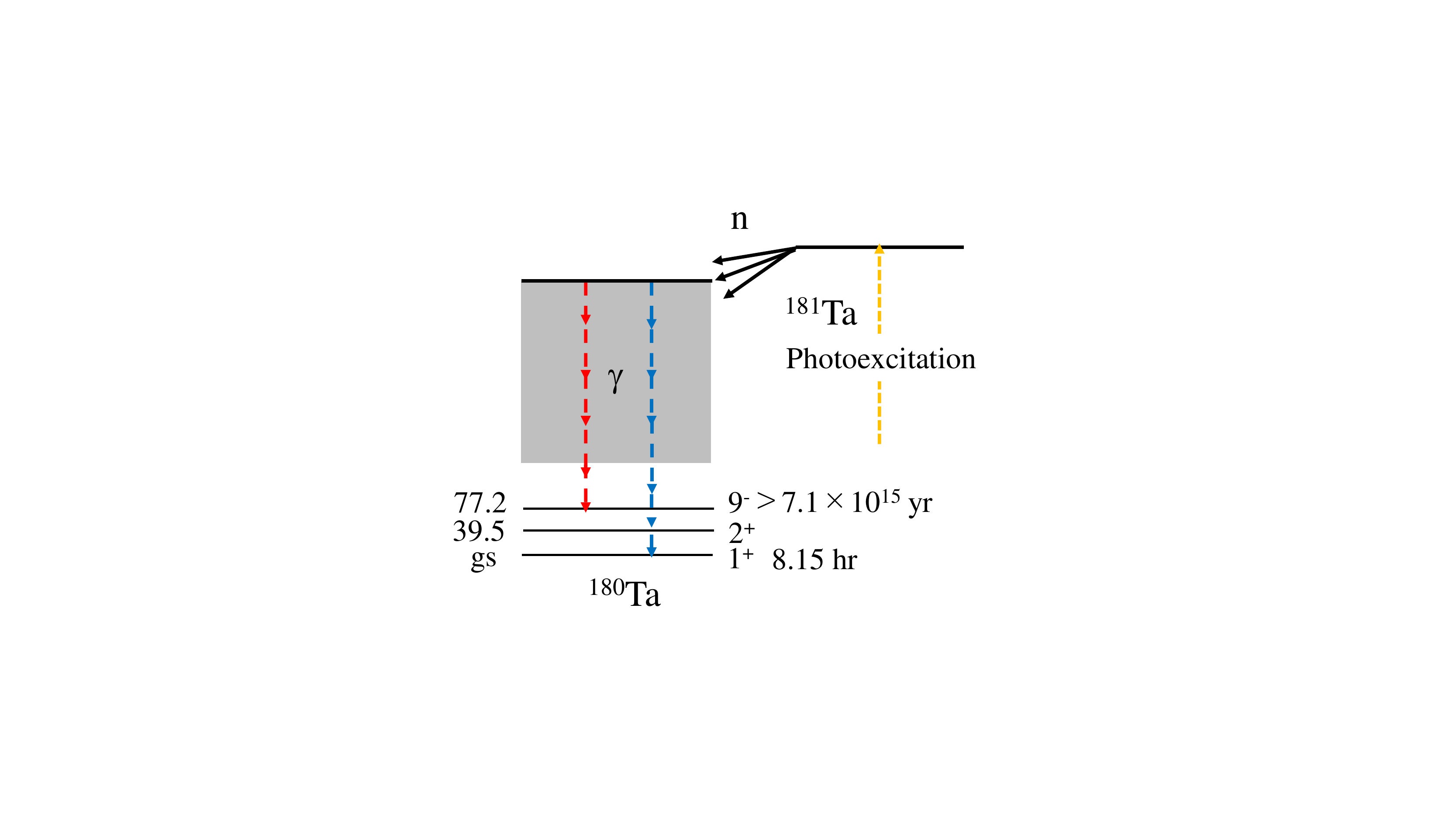}
\vskip -1cm
\caption{ (Color online) Illustration of the p-process production of $^{180}$Ta.  
}
\label{fig1}
\end{figure}

The $\gamma$-$\gamma$ coincidence experiment needs a neutron tag of the $^{181}$Ta($\gamma$,n)$^{180}$Ta reaction to identify mediating states in $^{180}$Ta that have access to both the isomeric and ground states. The neutron tagger can be either a moderator-based neutron detector or an assembly of liquid scintillation detectors.  We refer the p-process chaser detector with the former and latter taggers to as Type I and Type II, respectively. 
 
\subsection{p-process chaser detector Type I}
\begin{figure}
\center
\vskip -2cm
\includegraphics[bb = 100 0 700 420, width = 110 mm, clip]{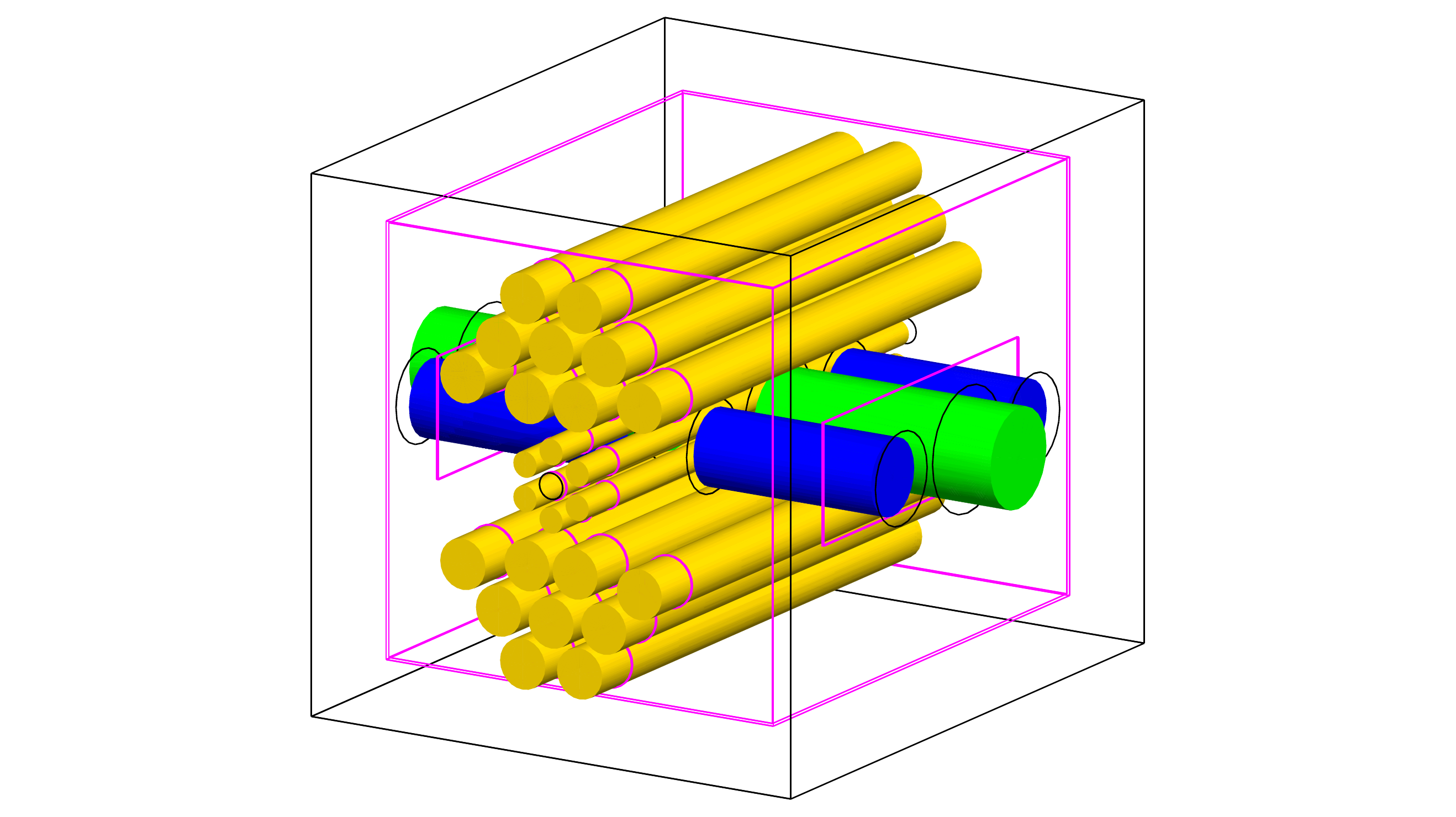}
\caption{ (Color online) Layout of the p-process chaser detector Type I.  The HPGe detectors are shown in green.  The LaBr$_3$(Ce) detectors are shown in blue.  The $^3$He counters are shown in yellow.
}
\label{fig2}
\end{figure}

Figure~\ref{fig2} depicts the p-process chaser detector Type I consisting of a moderator-based neutron detector, 2 high-purity Germanium (HPGe) detectors and 4 LaBr$_3$(Ce) detectors. Neutrons are detected with six $^3$He counters of 1-inch diameter and eighteen $^3$He counters of 2-inches diameter embedded in a polyethylene moderator.  The $^3$He gas pressure is 2 atmosphere. The size of the moderator is 556 mm (width) $\times$ 556 mm (height) $\times$ 656 mm (length). The six 1" $^3$He counters form a concentric ring at 35 mm from the $\gamma$-ray beam axis. The eighteen 2" $^3$He counters are embedded symmetrically in the upper and lower body of the moderator. The moderator-based detector plays the role of a high-efficiency neutron tagger for ($\gamma$,n) reactions. 

Two HPGe detectors with the crystal size of 80 mm in diameter and 70 mm in length are inserted perpendicular to the beam axis from the middle of the neutron moderator. The distance from the target to the front face of the detector is 62 mm. These HPGe detectors are used to identify $\gamma$ transitions between resonant states in $^{180}$Ta in $\gamma$-$\gamma$ coincidences.  A pair of LaBr$_3$(Ce) detectors with the crystal size of 3" in diameter and 4" in length sandwiches each of the HPGe detectors as shown in Fig. ~\ref{fig2}. The 4 LaBr$_3$(Ce) detectors are placed at 62 mm perpendicular to the beam axis.  The main purpose of these LaBr$_3$(Ce) detectors is to detect $\gamma$ transitions between unresolved states in the high NLD domain in coincidence with resonant $\gamma$ transitions leading to the isomeric and ground states detected with the HPGe detectors. 

\begin{figure}
\center
\includegraphics[bb = 0 100 600 600, width = 110 mm, clip]{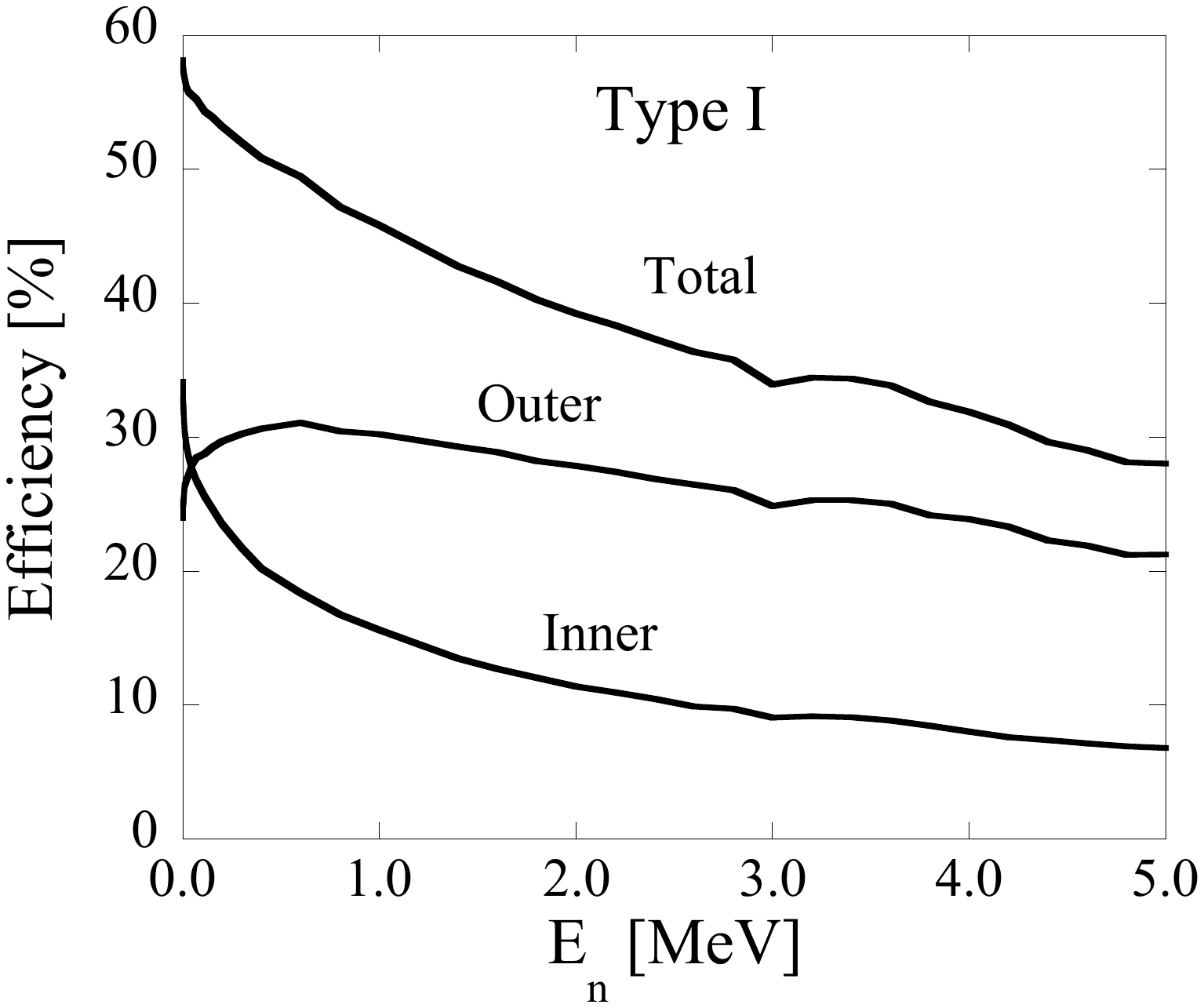}
\vskip -1.5cm
\caption{ (Color online) Neutron detection efficiencies of the p-process chaser detector Type I. The total efficiency and the efficiencies for the inner and outer $^3$He counters are shown.     
}
\label{fig3}
\end{figure}

\begin{figure}
\center
\vskip -3cm
\includegraphics[bb = 0 100 600 600, width = 110 mm, clip]{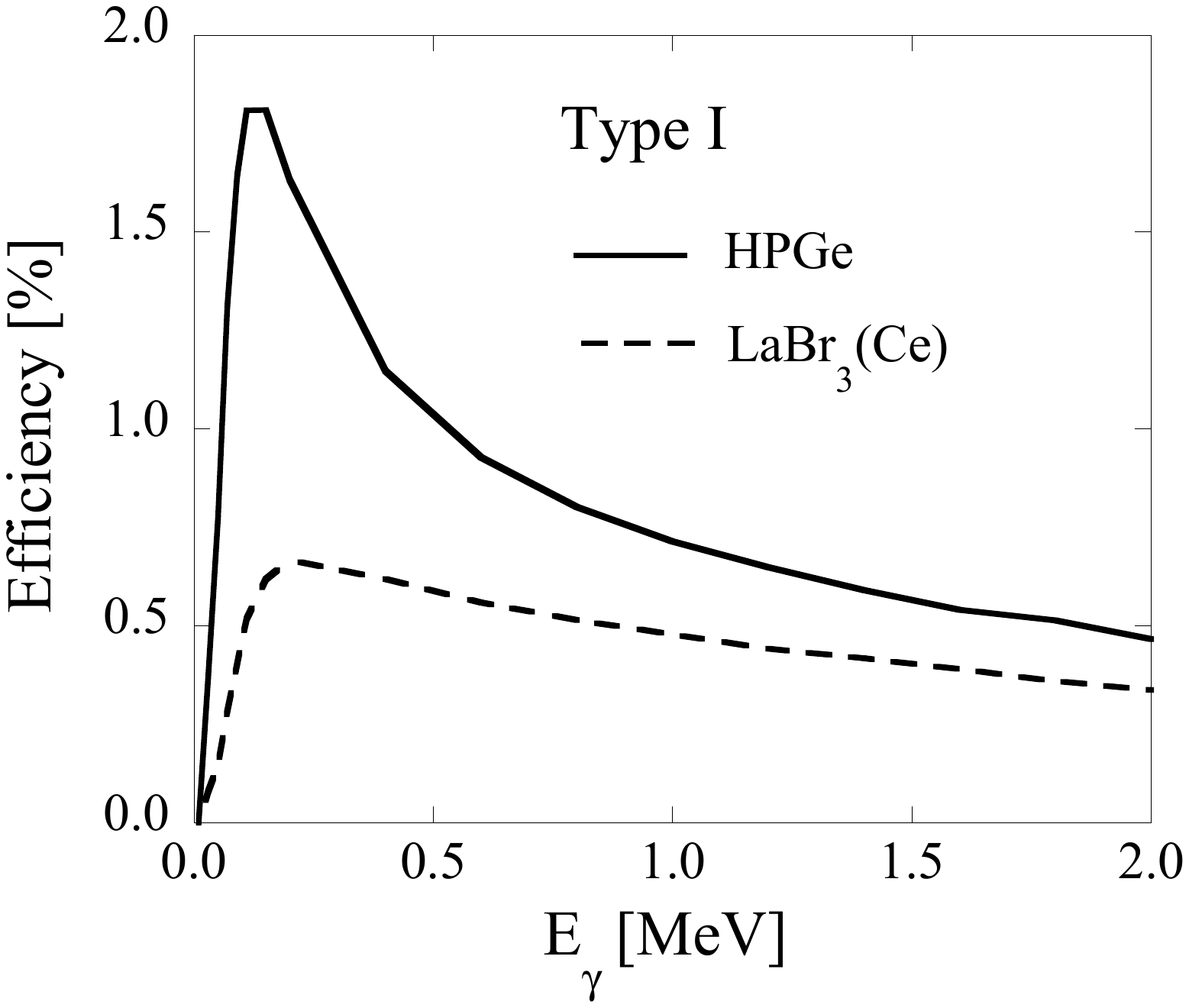}
\vskip -1.5cm
\caption{ (Color online) $\gamma$-ray detection efficiencies of the p-process chaser detector Type I. The efficiencies for each of the HPGe and LaBr$_3$(Ce) detectors are shown.}
\label{fig4}
\end{figure}

Geant4 simulations \cite{Alli06} were performed for the detector configuration of Type I shown in Fig. ~\ref{fig2}.  Figure~\ref{fig3} shows the total neutron detection efficiency as a function of the neutron energy along with its breakdown into efficiencies of the inner concentric ring of six $^3$He counters and outer eighteen $^3$He counters. Type I detector has no detection threshold for neutrons including thermal neutrons.  Figure~\ref{fig4} shows the photopeak efficiencies of the HPGe and LaBr$_3$(Ce) detectors as a function of the $\gamma$-energy.  

\subsection{p-process chaser detector Type II}
\begin{figure}
\center
\vskip -2cm
\includegraphics[bb = 100 0 700 420, width = 110 mm, clip]{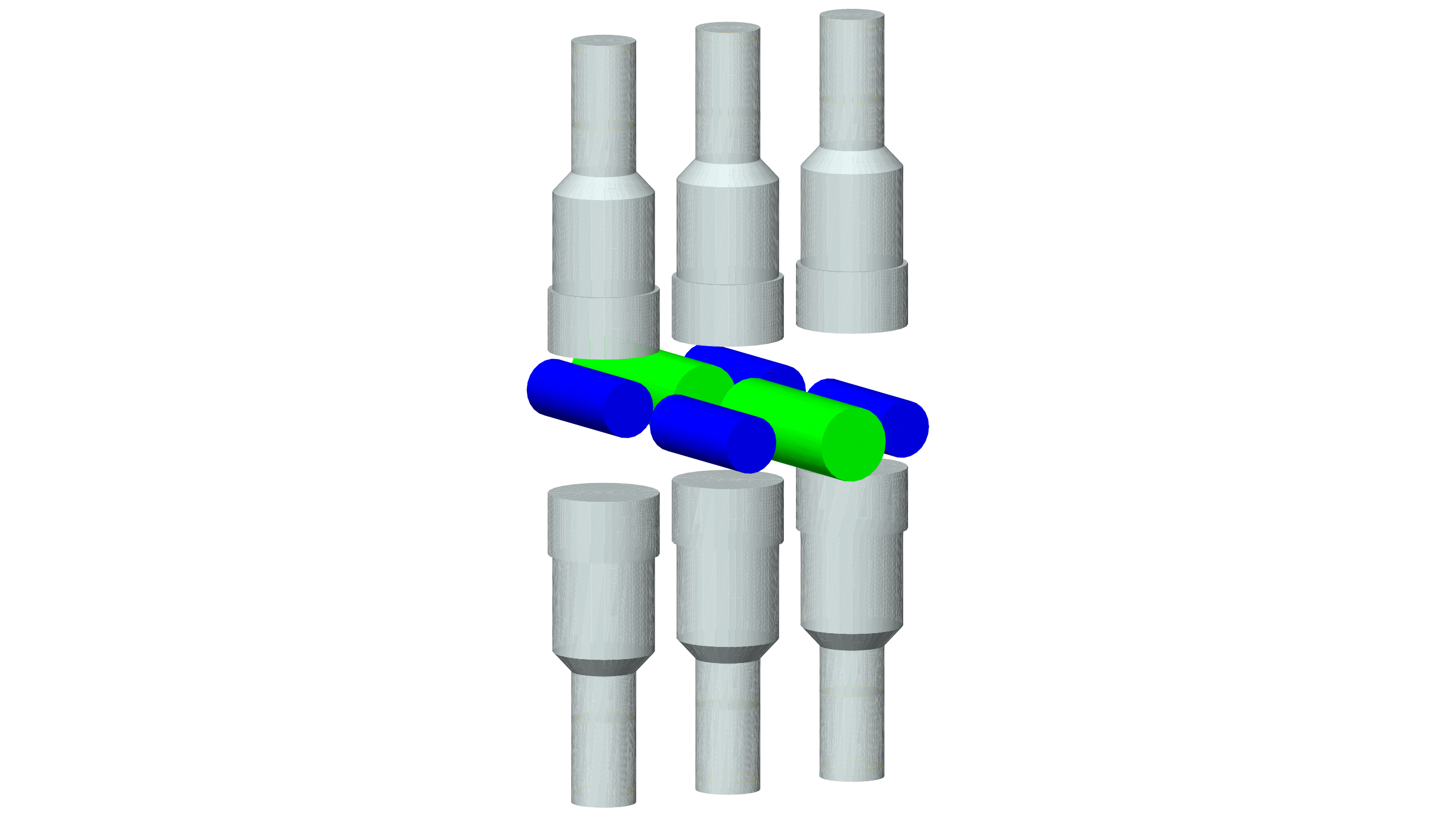}
\caption{ (Color online) Layout of the p-process chaser detector Type II.  The HPGe detectors are shown in green.  The LaBr$_3$(Ce) detectors are shown in blue.  The EJ-301 detectors are shown in grey.
}
\label{fig5}
\end{figure}

Figure~\ref{fig5} depicts the p-process chaser detector Type II consisting of an assembly of six EJ-301 liquid scintillation detectors, 2 high-purity Germanium (HPGe) detectors and 4 LaBr$_3$(Ce) detectors.  Three EJ-301 detectors of 5" in diameter and 2" in length are placed above the $\gamma$-ray beam axis and three more of the same size are placed below. The distance of the EJ-301 detectors from the $\gamma$-ray beam axis is 10 cm.  The configuration of the $\gamma$-ray detectors is the same for the chaser detector Type I. 
  
\begin{figure}
\center
\includegraphics[bb = 0 100 600 600, width = 110 mm, clip]{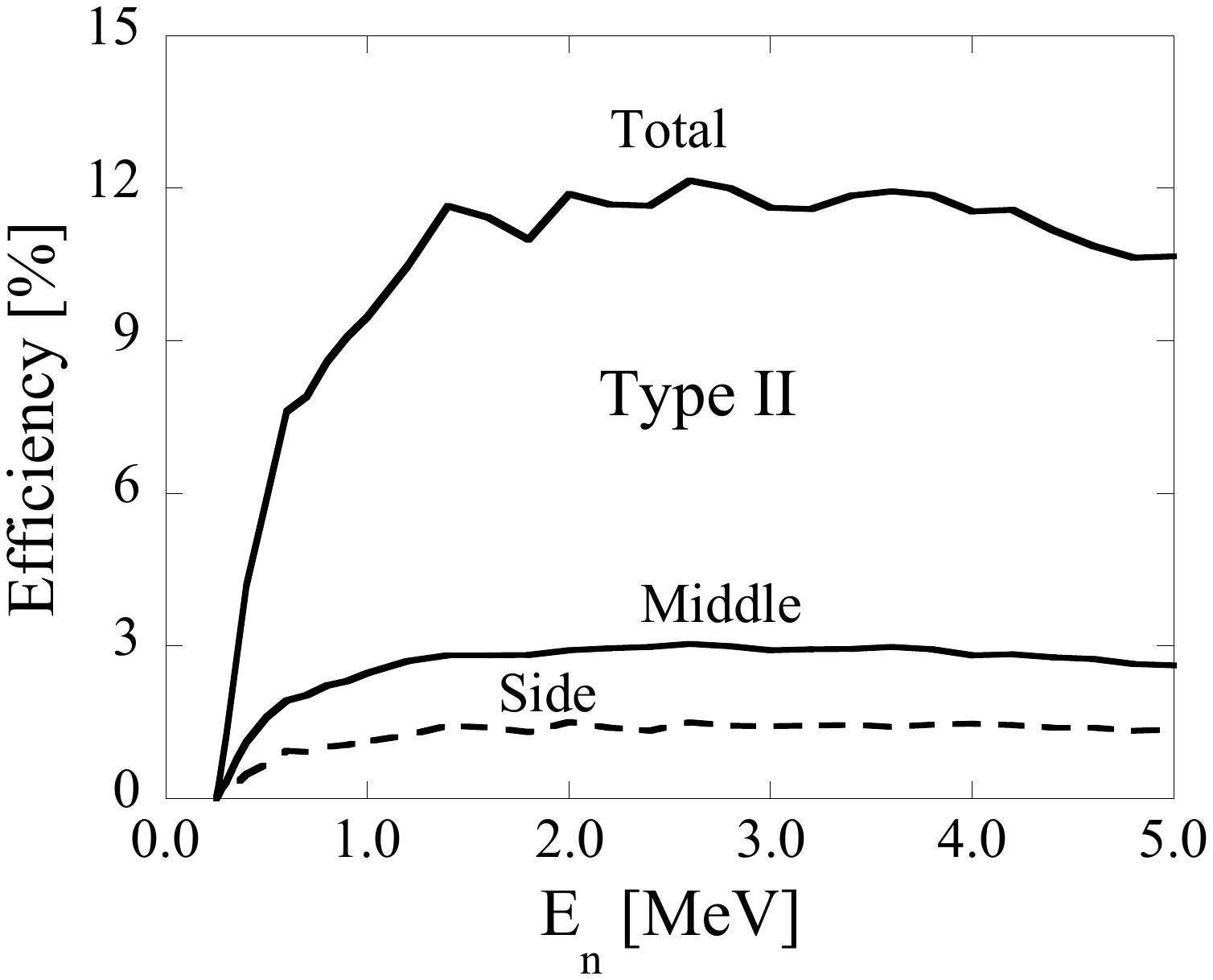}
\vskip -1.5cm
\caption{ (Color online) Neutron detection efficiencies of the p-process chaser detector Type II.  The total efficiency and the efficiencies for the EJ-301 detectors at the middle and side positions are shown.  
}
\label{fig6}
\end{figure}

\begin{figure}
\center
\includegraphics[bb = 0 100 600 600, width = 110 mm, clip]{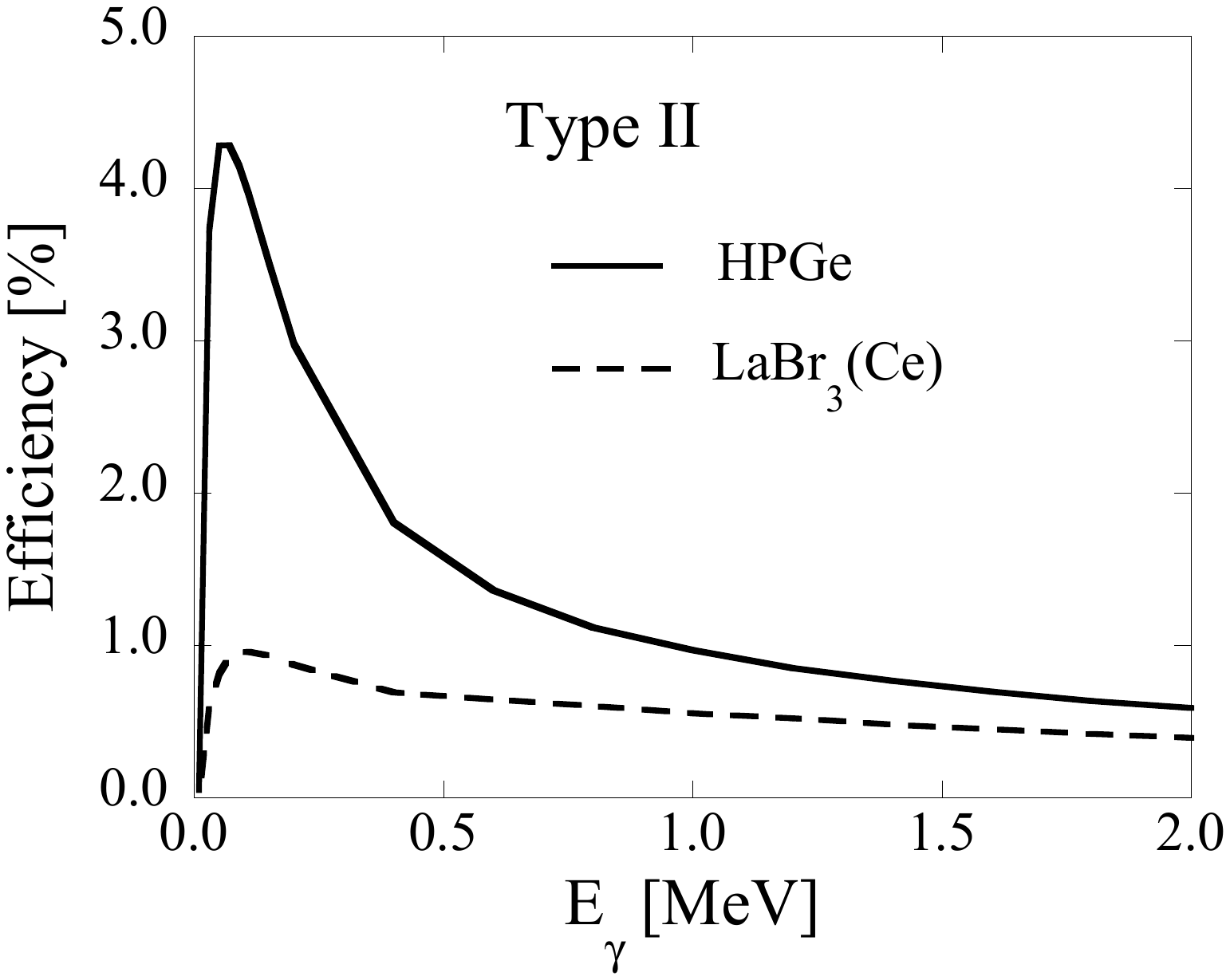}
\vskip -1.5cm
\caption{ (Color online) $\gamma$-ray detection efficiencies of the p-process chaser detector Type II. The efficiencies for each of the HPGe and LaBr$_3$(Ce) detectors are shown.
}
\label{fig7}
\end{figure}

Geant4 simulations were performed for the detector configuration of Type II shown in  Figure~\ref{fig5}. 
Figure~\ref{fig6} shows the neutron detection efficiency of the EJ-301 detector as a function of the neutron energy.  Liquid scintillation detectors are known to have the intrinsic detection efficiency at a discrimination level corresponding to a certain equivalent electron energy \cite{Lau93}.  The neutron detection efficiency in Fig.~\ref{fig6} was obtained for the equivalent electron energy (ee) of 30 keV \cite{Cav13} which is $\sim$ 250 keV in neutron energy. Figure~\ref{fig7} shows the photopeak efficiency of the HPGe and LaBr$_3$(Ce) detectors. The photopeak efficiency of the Type II detector is larger than that of the Type I detector because of the absence of the intervening polyethylene moderator and $^3$He counters.   

The neutron detection efficiency of the EJ-301 detector decreases with increasing the distance from a target.  The efficiency, for example at 1 MeV, decreases to 0.046 - 0.020 \% at the distance of 100 - 150 cm which is suited to a TOF measurement of neutrons. The neutron TOF measurement with this small efficiency is incompatible with the $\gamma_1$-$\gamma_2$ coincidence measurement. 

\section{Detection of mediating states along the p-process path}

The photoexcitation takes place nearly along the excitation energy (E) axis with $\Delta$J=1 in the E (excitation energy) - J (angular momentum) diagram unlike heavy-ion induced reactions used in the $\gamma$-spectroscopy experiment \cite{Drac98} by Dracoulis et al. The E1 photoexcitation of $^{181}$Ta with J$^{\pi}$ =7/2$^+$ populates states in $^{181}$Ta with 9/2$^-$, 7/2$^-$, and/or 5/2$^-$. The populated states dominantly undergo s-wave neutron decay to states in $^{180}$Ta with 5$^-$, 4$^-$, 3$^-$, and/or 2$^-$. Neutron decays with higher partial waves are hindered by the centrifugal potential. Thus, $\gamma$-transition starts from these low-spin states in $^{180}$Ta to the isomeric 9$^-$ state and ground 1$^+$ state. The $\gamma$-transition to the 9$^-$ state proceeds very selectively in cascade and progressively in overall ascending order of spin. 
In a naive picture of the E1 transition, the 9$^-$ state can be reached from 5$^-$ by transitions of 5$^-$ $\rightarrow$ 6$^+$ $\rightarrow$  7$^-$ $\rightarrow$ 8$^+$ $\rightarrow$ 9$^-$. 
The $\gamma$-transition to the ground state also selectively proceeds in cascade and progressively in overall descending order of spin. 

At mediating states, the cascade transition to the isomeric state is considered to branch to the transition path to the ground state. We propose to perform high-resolution $\gamma_1$-$\gamma_2$ coincidence measurements with the HPGe and LaBr$_3$(Ce) detectors following the p-process path. In this measurement, $\gamma_1$-$\gamma_2$ coincidences between the known last-step transitions to the isomeric and ground states \cite{NNDC,Drac98}, that is, between the 8$^+$ $\rightarrow$ 9$^-$ 100 keV or 10$^-$ $\rightarrow$ 9$^-$ 203 keV transition and 2$^+$ $\rightarrow$ 1$^+$ 40 keV or 0$^-$ $\rightarrow$ 1$^+$ 108 keV transition would provide experimental evidence for the presence of the mediating states.  
For discrete mediating states, reconstruction of the $\gamma$-transition scheme in $^{180}$Ta is made to identify the mediating state based on the experimental information on the well-known $\gamma$-transitions of the cascade provided by the $\gamma$-spectroscopic study \cite{Drac98}. The consistency of the excitation energy of the mediating state from the ground state and isomeric state is confirmed in the reconstruction.  In contrast, $\gamma$-transitions between unresolved mediating states in the high NLD domain may form a continuous energy distribution.  The continuous component for unresolved states can be extracted by unfolding the coincident $\gamma$-ray spectra with the response function of the $\gamma$ detectors.

\section{Data acquisition}

\subsection{p-process chaser detector Type I}
\begin{figure}
\center
\includegraphics[bb = 0 100 800 500, width = 110 mm, clip]{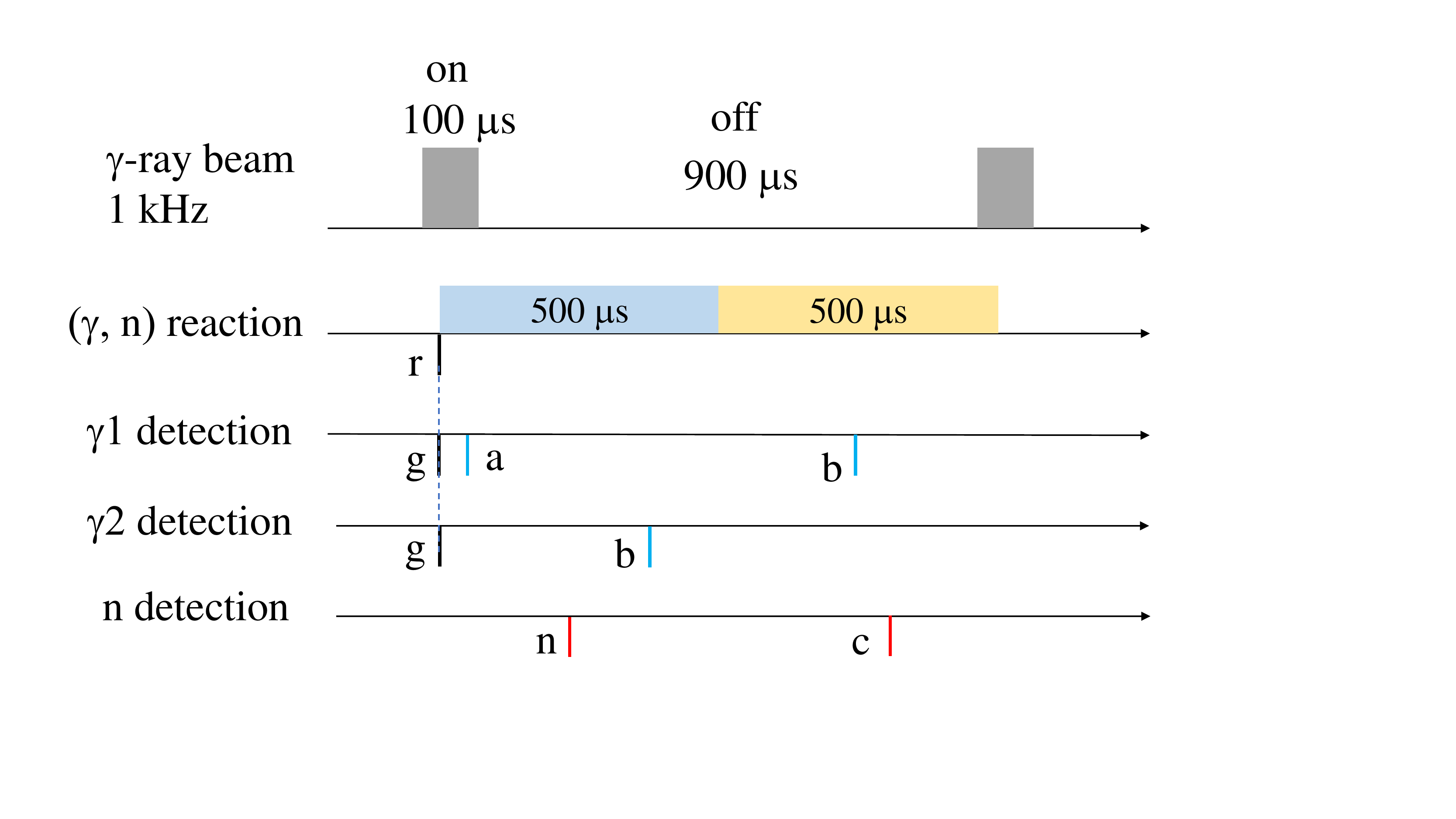}
\caption{ (Color online) Illustration of the time sequence of the data acquisition.  The $\gamma$-ray beam is produced at 1 kHz with the pulse width 100 $\mu$s.  Shown by the symbols are the $^{181}$Ta($\gamma,n)$ reaction event  ($r$), prompt $\gamma$ rays detected in coincidences ($g$), reaction neutron ($n$), background $\gamma$ event during the beam-on ($a$) and beam-off ($b$), and background neutron event ($c$).  See text for details. }
\label{fig8}
\end{figure}

Neutrons are detected after being moderated in the polyethylene moderator. The moderation time is as long as 500 $\mu$s \cite{Utsu17}.  In contrast, $\gamma$ rays are emitted promptly on the time scale of the ($\gamma$,n) reaction unless being emitted from isomeric states. The data acquisition is triggered by prompt $\gamma$ rays and stopped by the detection of moderated neutrons in the 1 ms time range every by event. 
Figure~\ref{fig8} illustrates the time scheme of the data acquisition.  In the SLEGS beamline of the Shanghai Light Source, $\gamma$ rays are produced in the laser Compton slant-scattering of photons from the 100 W CO$_2$ laser with 3.5 GeV electrons in the storage ring. A $\gamma$-ray beam is produced by operating the CO$_2$ laser at 1 kHz with the pulse width of 100 $\mu$s and laser power 10 W (10 \% duty cycle). 

We request the presence of two prompt $\gamma$ rays with the same time stamp during the beam-on (shown by the symbol $g$ in Fig. 8) to suppress background $\gamma$ events ($a$).  $\gamma$ rays detected during the beam-off are background $\gamma$ rays ($b$).  The reaction neutron from the ($\gamma$,n) reaction and background neutron are detected during the first 500 $\mu$s from the prompt $\gamma$ trigger ($n$), while only the background neutron is detected during the second 500 $\mu$s ($c$). 

\subsection{p-process chaser detector Type II}
The time correlation in detecting prompt neutrons and $\gamma$ rays is strong compared to that in detecting prompt $\gamma$ rays and moderated neutrons.  Therefore, a $\gamma$-ray beam produced by the operation of the laser at 100 \% duty cycle (100 W in SLEGS)  can be used in an experiment with the Type II detector, which is great advantage over an experiment with the Type I detector.           

\subsection{Background $\gamma$ events}
The gamma-ray beam produced in the laser Compton scattering is accompanied by synchrotron/bremsstrahlung radiation produced by an electron beam in the storage ring. The main component of the background radiation is keV X rays. Fine pencil-like gamma-ray beams with the background radiation are produced by the collimator system \cite{ZHao21}.
Background gamma events are produced in the atomic interaction (Compton scattering) of the pencil-like beam hitting the 100 $\mu$m target foil. The background event rate is estimated by a Geant4 simulation to be less than 1 k counts/s for the HPGe detector.  The contribution from the atomic interaction of high-energy gamma rays is negligible because of the forward-peaking angular distribution.  

\section{Event rate}
Let us estimate the event rate for the $n-\gamma_1-\gamma_2$ triple coincidence experiment with the p-process chaser detector.  Following the three steps of the p-process production of $^{180}$Ta, namely, the photoexcitation of $^{181}$Ta by a quasi-monochromatic $\gamma$-ray beam, neutron decay of $^{181}$Ta, and $\gamma$ decay of $^{180}$Ta as depicted in Fig. 1, the event rate $E$ is expressed by
\begin{equation}
R = n_{\gamma} \cdot n_t \cdot \sigma \cdot b \cdot _mC_2 \cdot \varepsilon_n \cdot \varepsilon_1 \cdot \varepsilon_2.
\label{eq:Event}
\end{equation}
Here $n_{\gamma}$ is the flux of the $\gamma$-ray beam, $n_t$ is the areal density of target nuclei,  $\sigma$ is the photoexcitation cross section, $b$ is the branching ratio of the neutron decay, $m$ is the number of $\gamma$ transitions ($\gamma$-ray multiplicity),  $\varepsilon_n$ is the neutron detection efficiency,  and $\varepsilon_i$ ($i$ = 1,2) is the $\gamma$-ray detection efficiency. In Eq.~(\ref{eq:Event}), we treat one neutron branching to an excited state in $^{180}$Ta followed by the detection of two $\gamma$ rays, $\gamma_1$ and $\gamma_2$, out of $m$ $\gamma$ rays in coincidence as represented by $_m$C$_2$=$m(m-1)$/2. When one performs an experiment with the chaser detector of Type I or Type II at 10 MeV with a flux of 10$^5$ photons/s \cite{ZHao21,HXu22} and a Ta foil of 100 $\mu$m thickness under the conditions listed in Table 1, one finds the event rate summarized in Table 2. 
In this estimate, we assumed for simplicity that the neutron energy $E_n$=1 MeV, the $\gamma$-ray energy E$_{\gamma}$=200 keV, the branching ratio $b$= 0.2, and the neutron multiplicity $m$=7.  In Table 2, the rate for $b$=1 is given because the neutron decay occurs at 100 \% of the time.  We assumed to use a thin target foil to avoid absorption of low-energy $\gamma$ rays in the target. It is to be noted that the event rate can be increased by using multi-target foils placed along the $\gamma$-ray beam axis.   

\begin{table}
\label{tab1}
\center
\caption{Experimental model parameters}
\begin{tabular}{ccccccc}
\hline 
\hline 
$n_{\gamma}$ & $n_t$ &  $\sigma$ & $b$ & $m$ & $E_n$ & $E_{\gamma}$ \\ 
(photons/s) &  (nuclei/cm$^2$) & (mb) & & & (MeV) & (keV) \\
\hline
10$^5$  & 5.5 $\times$ 10$^{20}$ & 80  & 0.2 & 7 & 1 & 200 \\
\hline
\hline
\end{tabular}
\end{table}

\begin{table}
\label{tab2}
\center
\caption{Detection efficiencies and event rates of Type I and Type II chaser detectors for $b$=1. The experimental model parameters given in Table I are used except $b$. }
\begin{tabular}{cccccc}
\hline
\hline
 & $\varepsilon_n$  & $\varepsilon(HPGe)$ & $\varepsilon^{\ast}(LaBr)$ & R(HPGe-HPGe) & R(HPGe-LaBr) \\
 &  & & & (counts/hr) & (counts/hr)  \\
\hline
Type I & 0.46 & 0.016 & 0.027 & 41 & 67 \\
Type II & 0.095 & 0.030 & 0.035 & 28 & 33 \\
\hline
\hline
\end{tabular}
$^*$ Sum efficiency of four LaBr$_3$(Ce) detectors
\end{table}

The partial cross section measurement showed that about 10 \% of the total $^{181}$Ta($\gamma$,n)$^{180}$Ta cross section is attributed to the isomeric state \cite{Goko06}.  Therefore, the event rate of the $n$-$\gamma_1$-$\gamma_2$ triple coincidence measurement reaching the isomeric and ground states are $\sim$ 10 \% and $\sim$ 90 \%, respectively, of the rate given in Table 2.  

\section{Summary}
The presence of mediating states in $^{180}$Ta through which the long-lived isomeric state $^{180}$Ta$^m$ and short-lived ground state $^{180}$Ta$^{gs}$ are thermalized in the p-process nucleosynthesis has remained mysterious. 
A best possible way to identify mediating states is to directly chase $\gamma$ transitions to the isomeric and ground states following the neutron emission in the $^{181}$Ta($\gamma$,n)$^{180}$Ta reaction. Two types of the p-process chaser detectors (not constructed) are proposed based on $n$-$\gamma_1$-$\gamma_2$ triple coincidences with different neutron tagging detectors, a moderator-based neutron detector (Type I) and an assembly of liquid scintillation detectors (Type II). 
With these chaser detectors, one can search for resonant states and unresolved states in the high NLD domain as mediating states.  

In general, the proposed p-process chaser detectors help to investigate the $\gamma$-transition scheme unique to the nucleosynthesis of p-nuclei.    

\section{Acknowledgements}
This work was supported by the Chinese Academy of Sciences President's International Fellowship Initiative under Grant No. 2021VMA0025, the National Natural Science Foundation of China under Grant No. 11875311 and the Strategic Priority Research Program of Chinese Academy of Sciences under Grant No. XDB34030000.  


\begin{thebibliography}{}
\bibitem{ArGo03} M.  Arnould and S. Goriely, The p-process of stellar nucleosynthesis: Astrophysics and nuclear physics status, Physics Reports 382, (2003) 1.
\bibitem{WH78} S.E. Woosley and W.M. Howard, THE p-PROCESS IN SUPERNOVAE, Astrophys. J. Suppl. 36, (1978) 285.
\bibitem{Ray95} M. Rayet et al., The p-process of Type II supernovae, Astro. Astrophys. 298 (1995) 517. 
\bibitem{Rau02} T. Rauscher et al., NUCLEOSYNTHESIS IN MASSIVE STARS WITH IMPROVED NUCLEAR AND STELLAR PHYSICS, Astrophys. J. 576 (2002) 323.
\bibitem{Utsu06} H. Utsunomiya et al., Direct determination of photodisintegration cross sections and the p-process, Nucl. Phys. A 777 (2006) 459.
\bibitem{Trav15} C. Travaglio et al., Testing the role of SNe Ia for galactic chemical evolution of p-nuclei with two-dimensional models and with s-process seeds at different metallicities, Astrophys. J. 799 (2015) 54. 
\bibitem{Ande89} E. Anders and N. Grevesse, Abundances of the elements: Meteoritic and solar, Geochim. Cosmochim. Acta 53 (1989) 197. 
\bibitem{Palm93} H. Palm and H. Beer in: H.H. Vogt (Ed.), Landolt B\"ornstein, New Series, Group VI, Astro. Astrophys. Subvolume 3a (1993) 196. 
\bibitem{NNDC} https://www.nndc.bnl.gov/nudat3/chartNuc.jsp
\bibitem{Utsu03} H. Utsunomiya et al., Cross section measurements of the $^{181}$Ta($\gamma$,n)$^{180}$Ta reaction near neutron threshold and the p-process nucleosynthesis, Phys. Rev. C 67 (20030 015807.
\bibitem{Drac98} G.D. Dracoulis et al., Intrinsic states and collective structure in $^{180}$Ta, Phys. Rev. C 58 (1998) 1444.
\bibitem{Beli99} D. Belic et al., Photoactivation of $^{180}$Ta$^m$ and its implications for the nucleosynthesis of nature's rarest naturally occurring isotope, Phys. Rev. Lett. 83 (1999) 5242. 
\bibitem{Goko06} S. Goko et al., Partial photoneutron cross sections for the isomeric state $^{180}$Ta$^m$, Phys. Rev. Lett. 96 (2006) 192501. 
\bibitem{YoTa83} K. Yokoi and K. Takahashi, Slow neutron capture origin for $^{180}$Ta$^m$, Nature (London) 305 (1983) 198.
\bibitem{Hila01} S. Hilaire, J.P. Delaroche, M. Girod, Combinatorial nuclear level densities based on the Gogny nucleon-nucleon effective interaction, Eur. Phys. J. A 12 (2001) 169.
\bibitem{Gori08} S. Goriely, S. Hilaire, A.J. Koning, Improved microscopic nuclear level densities within the Hartree-Fock-Bogoliubov plus combinatorial method, Phys. Rev. C 78 (2008) 064307. 
\bibitem{Cous84} T. Cousins et al., Population of $^{180}$Ta states via the monochromatic reaction $^{181}$Ta($\gamma$, n)$^{180}$Ta, Phys. Rev. C 29 (1984) 1085. 
\bibitem{Alli06} J. Allison et al., Geant4 developments and applications, IEEE Trans. Nucl. Sci. 53 (2006) 270. 
\bibitem{Lau93} H. Laurent et al., EDEN: a neutron time-of-flight multidetector for decay studies of giant states, Nucl. Inst. and Methods in Physics Research A 326 (1993) 517. 
\bibitem{Cav13} M. Cavallaro et al., Pulse-shape discrimination in NE213 liquid scintillator detectors, Nucl. Inst. and Methods in Physics Research A 700 (2013) 65. 
\bibitem{Utsu17} H. Utsunomiya et al., Direct neutron multiplicity sorting with a flat-efficiency detector, Nucl. Inst. and Methods in Physics Research A 871 (2017) 135. 
\bibitem{ZHao21} Z.R. Hao et al., Collimator system of SLEGS beamline at Shanghai Light Source, Nucl. Inst. and Methods in Physics Research A 1013 (2021) 165636.
\bibitem{HXu22} H.H. Xu et al., Interaction chamber for laser Compton slant-scattering in SLEGS beamline at Shanghai Light Source, Nucl. Inst. and Methods in Physics Research A, submitted. 
\end{thebibliography}

\end{document}